\documentclass{elsart}
\usepackage{graphicx,amssymb}
\usepackage[english]{babel}

\newcommand{\debhu}{Debye-H\"{u}ckel}

\begin{document}

%%%%%%%%%%%%%%%%%%%%%%%%%%%%%%%%%%%%%%%%%%%%%%%%%%%%%%%%%%%%%%%%%%%%%%%%%%%%%%%%%%%%%%%%%%%%%%%%%%%%%%%%%%%%%%%%
\begin{frontmatter}

\title{Metastable and stable equilibrium states of stellar electron-nuclear plasmas}

\author[Poli,INFNTo]{F.~Ferro},
\author[Poli,INFNTo]{A.~Lavagno},
\author[Poli,INFNCa]{P.~Quarati}
\address[Poli]{Dipartimento di Fisica, Politecnico di Torino, I-10129 Torino, Italy}
\address[INFNTo]{INFN - Sezione di Torino, I-10125 Torino, Italy}
\address[INFNCa]{INFN - Sezione di Cagliari, I-09042 Monserrato, Italy}

\date{27 January 2004}

\begin{abstract}
By minimizing free energy density, we show that the stellar core of a hydrogen burning star is not in a global
thermodynamical equilibrium unless density, temperature, mass and composition assume given values. The core (as the
solar interior) may be viewed more appropriately as a metastable state with very long lifetime. Slightly non-extensive
distribution function could be the natural distribution for a weakly non-ideal plasma like a stellar core and represents a
more appropriate approximation to this system than a Maxwellian distribution, 
without affecting bulk properties of stars.
\end{abstract}


\end{frontmatter}
%%%%%%%%%%%%%%%%%%%%%%%%%%%%%%%%%%%%%%%%%%%%%%%%%%%%%%%%%%%%%%%%%%%%%%%%%%%%%%%%%%%%%%%%%%%%%%%%%%%%%%%%%%%%%%%%

%%%%%%%%%%%%%%%%%%%%%%
\section{Introduction}
%%%%%%%%%%%%%%%%%%%%%%
Landau and Lif\v{s}its~\cite{landaubook} distinguish between two different equilibrium states called metastable and
stable states. A system in a metastable state cannot return to its initial state after a sufficiently large deviation
from it and could pass to another (meta)stable state. Although metastable states are stable within certain limits,
after a time that could be very long, the system will pass to another state. A system removed from the stable state
with maximum entropy will sooner or later return to it.

Sewell~\cite{sewell} shows that the equilibrium states are globally thermodynamically stable if they minimize the free
energy density of the system, satisfy KMS (Kubo-Martin-Schwinger) fluctuation-dissipation conditions and are of
Gibbsian type. Metastable states (ideal with infinite lifetime or normal with finite long lifetime) have a minimum of
free energy density only in some reduced space state and these states are always supported only by models
with many-body, long-range interactions among the particles (as, for instance, gravitational, Coulomb, spin-glass
interactions). Considering infinite systems, Sewell minimizes {\it density} of free energy to avoid divergences. The
minimization of this quantity allows us to analyze local, finite subsystems of the entire system (like, for instance,
as small as the \debhu\/ plasma sphere).

Recently, it has been shown that metastable states can naturally be described by distributions accounting for
temperature fluctuations~\cite{wilk,beck} or by quasi-particle models~\cite{CLMQ}, and belonging to the generalized
statistical mechanics that now is usually called non-extensive thermostatistics~\cite{KLR}.

In this work we are interested in non-relativistic, hot, dense electron-nuclear plasmas (like the core of stars or the
solar interior) and discuss the conditions required for their equilibrium or their metaequilibrium (metastability), following the approach outlined by Landau and Lif\v{s}its~\cite{landaubook2} and recently applied by
Vasiliev~\cite{vasiliev}, but minimizing free energy {\em density\/} and total energy {\em density\/}.

It is commonly accepted that stars like the Sun have a core, i.e. an electron-nuclear plasma, in equilibrium
conditions, that can be described by an electron and an ion velocity distribution which are Maxwellian 
(local thermodynamical equilibrium).

Main sequence stars produce a luminosity equal to the heat production rate and reach a stationary regime. Nuclear
fusion reactions produce the heat and do not sensibly perturb the state of the Sun. This stationary state that is
usually called, in astrophysics, equilibrium state, has a long life-time, until the nuclear fuel is completely burned
out. Stars are not in thermal equilibrium at present because they radiate and produce heat in the core with a
temperature decreasing from core to surface. Thermal exchanges take a very long time, whereas gravitational
equilibrium is rapidly reached. Nuclear reactions are so slow that in the time interval between one fusion reaction
and another the system returns to the Maxwellian equilibrium due to many binary Coulomb collisions among the
ions~\cite{bahcallbook}.

We want to point out that the return to the initial state after a fusion reaction is also possible if the distribution
function describing the initial state is a stationary (stable or metastable) one, analytically expressed by a deformed
Maxwellian (e.g. a Druyvenstein distribution) or a non-extensive distribution. What counts to decide the formal
expression of the distribution is the type of collisions in which the ions are involved and the dependence on momentum
of the elastic collisional cross sections (Coulomb, screened Coulomb, enforced Coulomb, among others), or the complete
absence of ion-ion correlation and interaction \cite{newferro}. Therefore, this argument based on the slowness of fusion reactions is
not useful to decide whether the system (the electron-nuclear core) is really in an equilibrium state (globally
thermodynamically stable) or is close to it, or rather it is a metastable state, and to decide what distribution is
the best approximation.

In this work, we deduce the values of temperature, density and core radius of stellar interiors having a given
composition at which they are in equilibrium (or in metaequilibrium state), supposing stellar cores made of ideal
ionized gas plus corrections due to particles identity and internuclear interactions, as indicated in
Refs.~\cite{landaubook2,vasiliev}.

We find that globally thermodynamically stable equilibrium is reached in stars of the same size and composition of the
Sun, at the core thermal energy of $\mathrm{k_B} T\approx 5\, \mathrm{keV}$ (more than three times higher than the
actual temperature of the solar interior), and at an electron density of $n\approx 0.027\cdot 10^{-12}\,
\mathrm{fm}^{-3}$ (about half the actual value in the solar core, that must have a volume of the central core of
radius $R\approx 0.2\, R_{\odot}$).

Therefore, Maxwellian distributions are of course very good approximations (zero approximations), considering the high
precision reached by SSM (Standard Solar Model) 
in describing solar neutrinos production. However a better approximation should be the
distribution of a metastable state, possibly given in an effort to accommodate equilibrium and metaequilibrium states, 
as recently developed in non-extensive thermostatistics, in spite of criticism devoted to its use in this
field~\cite{bahcall2,nauenberg}.

As already shown in the recent past, very small deviations from Maxwellian momentum distribution do not modify the
properties of stellar core, are in agreement with the helioseismology constraints, but may affect the evaluation of the
nuclear fusion rates that may be enhanced or depleted, depending on superdiffusion or subdiffusion property of the
particles~\cite{LavQua}. Let us mention that deformed distributions have been recently used also in the
interpretation of many heavy ions processes in thermalization and equilibration of quark-gluon plasma~\cite{alberico}; 
a study of big bang nucleosynthesis of lithium and of a possible solution of the problem of lithium abundance in astrophysical systems 
have been undertaken by us and is in progress, among other applications~\cite{KLR}.

%%%%%%%%%%%%%%%%%%%%%%%%%%%
\section{Metastable states}
%%%%%%%%%%%%%%%%%%%%%%%%%%%
Effort to accommodate a fundamental characterization of either equilibrium or metastable states in terms of certain
stability conditions was undertaken since long time. The equilibrium states are globally thermodynamically stable, and
are those that minimize the free energy density of the system and satisfy KMS conditions. Grades of metastability
correspond to different local stability properties.

Metastable states (ideal) do not minimize free energy density of the entire system, but minimize the restriction of
free energy density to some reduced state space. They are long-lived, supported by long-range, many-body forces.
They are locally thermodynamically stable if the free energy of the system cannot be lowered by modifications of the
state that is confined to bounded regions of space~\cite{sewell}.

The metastability of a state depends on the thermal perturbations to which the system is subjected. A state may have
very long lifetime in presence of certain thermal perturbations and may be short-lived when subjected to others. Of
course, in presence of perturbations or thermal fluctuations, the distribution function of the system differs more
or less from the Maxwellian distribution.

By using well known results of Landau and Lif\v{s}its~\cite{landaubook2}, and following the approach of
Vasiliev~\cite{vasiliev}, we examine the conditions for equilibrium by minimizing the free energy density. The free
energy of an ideal electron gas is
\begin{equation}
F_{\mathrm{ideal}}=-\mathrm{k_B} T N-\mathrm{k_B} T N\ln\left\{\frac{1}{n} \left[\frac{m\mathrm{c}^2\mathrm{k_B}
T}{2\pi(\hbar\mathrm{c})^2}\right]^{3/2}\right\}\;,  \label{Fid}
\end{equation}
where $N$ and $m$ are, respectively, the number and the mass of electrons, $n$ is the electron density,
$\mathrm{k_B}T$ is the thermal energy of the plasma.

Let us consider now a hot dense electron-nuclear plasma: the free energy should be written, starting from
Eq.~(\ref{Fid}), in the following form
\begin{equation}
F=F_{\mathrm{ideal}}+F_{\mathrm{quantum}}+F_{\mathrm{correlations}}\; ,\label{Factual}
\end{equation}
where $F_{\mathrm{quantum}}$ and $F_{\mathrm{correlations}}$ account for the main corrections that arise in real
gases, namely the quantum exclusion principle between fermions and correlation effects between charged particles.

The first correction, $F_{\mathrm{quantum}}$, can be written as
\begin{equation}
F_{\mathrm{quantum}}=+NE_1 n\; ,  \label{corr1}
\end{equation}
with
\begin{displaymath}
E_1=\left(\frac{\pi^{3/2}\mathrm{a}_0^{3/2}\e^3}{4}\right)\frac{1}{(\mathrm{k_B} T)^{1/2}}=\frac{e_1}
{(\mathrm{k_B}T)^{1/2}}\; ,
\end{displaymath}
where $\mathrm{a}_0\simeq 0.53\cdot 10^5\, \mathrm{fm}$ is the Bohr radius, and $e_1\simeq 2.68\cdot 10^7\,
\mathrm{MeV}^{3/2}\mathrm{fm}^3$.

The second correction is
\begin{equation}
F_{\mathrm{correlation}}=-NE_2 n^{1/2}\; ,  \label{corr2}
\end{equation}
with
\begin{displaymath}
E_2=\left(\frac{2\pi^{1/2}\e^3}{3}\right)\frac{(\bar{Z}+1)^{3/2}}{(\mathrm{k_B} T)^{1/2}}=
e_2\frac{(\bar{Z}+1)^{3/2}}{(\mathrm{k_B} T)^{1/2}}\; ,
\end{displaymath}
where $\bar{Z}$ is the average chemical composition and $e_2\simeq 2.04\,\mathrm{MeV}^{3/2}\mathrm{fm}^3$.

If we define
\begin{displaymath}
a=\left[\frac{m\mathrm{c}^2\mathrm{k_B} T}{2\pi(\hbar{\mathrm{c}})^2}\right]^{3/2}= a_1(\mathrm{k_B} T)^{3/2}\; ,
\end{displaymath}
where $a_1\simeq 0.28\cdot 10^{-8}\, \mathrm{MeV}^{-3/2}\mathrm{fm}^{-3}$, the free energy density, starting from
Eqs.(\ref{Factual}), (\ref{corr1}) and~(\ref{corr2}), is 
\begin{displaymath}
\frac{F}{V}=-\mathrm{k_B} T n-\mathrm{k_B} T n\ln\left(\frac{a}{n}\right)+E_1 n^2-E_2 n^{3/2}\; ,
\end{displaymath}
$V$ being the volume of the plasma.

Now, we should calculate the value $n_*$ of particle density that minimizes the $F/V$ free energy density: thus,
we can write the extremal condition
\begin{equation}
\left.\frac{\partial (F/V)}{\partial n}\right|_{T,N} =0\; ,
\label{minimum condition}
\end{equation}
from which we obtain the non-linear equation
\begin{equation}
\ln\left [ \frac{n}{a_1 (\mathrm{k_B} T)^{3/2}}\right ]+2\frac{e_1}{(\mathrm{k_B}T)^{3/2}}\,n-\frac{3}{2}
e_2\frac{(\bar{Z}+1)^{3/2}}{(\mathrm{k_B} T)^{3/2}}\,n^{1/2}=0\; ,  \label{condition1}
\end{equation}
that links together $T$, $n$ and the chemical composition $\bar{Z}$.

It can be shown that condition~(\ref{minimum condition}) is not only necessary, but also sufficient to assure that the
root of Eq.(\ref{condition1}) is the actual minimum of free energy density, provided that the inequality
\begin{displaymath}
\frac{(\bar{Z}+1)^3}{(\mathrm{k_B} T)^{3/2}}<9.17\cdot 10^7\, {\mathrm{MeV}^{-3/2}}\; ,
\end{displaymath}
is satisfied. Astrophysical plasmas of stellar cores usually have $\mathrm{k_B} T\sim 1\, \mathrm{keV}$ and $\bar{Z}$
of order unity: then the previous inequality holds for these systems.

Looking for a second relation linking $n$ and $T$, we can take the total energy of non-relativistic electron-nuclear
plasma as
\begin{equation}
E_{\mathrm{total}}=U+E_{\mathrm{kinetic}}+\frac{3}{2}NE_1 n-\frac{3}{2} NE_2 n^{1/2}+
\frac{\pi^2}{15}\left(\frac{\mathrm{k_B} T}{\hbar{\mathrm{c}}} \right)^3 V\mathrm{k_B} T\; ,  \label{Etotal}
\end{equation}
where the last term represents the black-body radiation contribution.

According to virial theorem~\cite{vasiliev virial}, the potential energy $U$ of particles with Coulomb interaction is
equal to their double kinetic energy with opposite sign, namely
\begin{equation}
U=-2\,\frac{3}{2} N\mathrm{k_B} T=-3N\mathrm{k_B} T\; .  \label{virial}
\end{equation}

According to Eqs.~(\ref{Etotal}) and~(\ref{virial}), we obtain the total energy density
\begin{displaymath}
\frac{E_{\mathrm{total}}}{V}=-\frac{3}{2}\mathrm{k_B} T n+\frac{3}{2}\frac{e_1}{(\mathrm{k_B} T)^{1/2}}
n^2-\frac{3}{2}\frac{e_2(\bar{Z}+1)^{3/2}}{(\mathrm{k_B} T)^{1/2}}n^{3/2}+\frac{\pi^2}{15} \frac{(\mathrm{k_B}
T)^4}{(\hbar\mathrm{c})^3}\; ,
\end{displaymath}
whose minimum condition,
\begin{displaymath}
\left.\frac{\partial (E_{\mathrm{total}}/V)}{\partial (\mathrm{k_B} T)}\right|_{N,V}=0\; ,
\end{displaymath}
can now be written as
\begin{equation}
-\frac{3}{2}n-\frac{3}{4}\frac{e_1}{(\mathrm{k_B} T)^{3/2}} n^2 +\frac{3}{4} \frac{e_2(\bar{Z}+1)^{3/2}}{(\mathrm{k_B}
T)^{3/2}}n^{3/2}+\frac{4\pi^2}{15} \left(\frac{\mathrm{k_B} T}{\hbar{\mathrm{c}}}\right)^3=0\; . \label{energy density
minimum}
\end{equation}

Eq.~(\ref{energy density minimum}) is the second equation that links $T$, $n$ and $\bar{Z}$ together.

Considering that, in a Sun-like star, $n\sim 10^{-14}\,\mathrm{fm}^{-3}$, $\mathrm{k_B} T\sim 10^{-3}\,\mathrm{MeV}$
and $\bar{Z}\sim 1.25$, we can neglect the terms with $e_1$ and $e_2$ in Eq.~(\ref{energy density minimum}), thus
obtaining
\begin{equation}
\mathrm{k_B} T=\left[\frac{3}{2}n\frac{15}{4\pi^2}(\hbar{\mathrm{c}})^3\right]^{1/3}= b^{1/3}n^{1/3}\; ,
\label{condition2}
\end{equation}
where $b\simeq 4.56\cdot 10^6\,\mathrm{MeV}^3\mathrm{fm}^3$.

Therefore, $\mathrm{k_B} T$ and $n$ must satisfy both Eq.~(\ref{condition1}) and Eq.~(\ref{condition2}), given a fixed
chemical composition, $\bar{Z}$. If we choose $\bar{Z}=1.25$, i.e. the Sun composition, we find out from the previous
discussion that the plasma is globally thermodynamically stable if and only if its electron density $n_*$ and its
thermal energy $\mathrm{k_B} T_*$ are, respectively,
\begin{equation}
n_*\simeq 2.74\cdot 10^{-14}\,\mathrm{fm}^{-3}\;\;\mathrm{and}\;\;\mathrm{k_B} T_*\simeq 5\cdot
10^{-3}\,\mathrm{MeV}\; .  \label{equilibrium values}
\end{equation}

Let us now consider a star characterized by the values of $n_*$ and $\mathrm{k_B} T_*$ given above in
Eq.~(\ref{equilibrium values}); we want to determine if its mass, $M_*$, could satisfy the condition $M_*\approx
M_{\odot}$, where $M_{\odot}\simeq 1.99\cdot 10^{33}\,\mathrm{g}$ is the solar mass. We suppose the star in
hydrostatic equilibrium, namely the gravitational pressure, which tends to compress the star, is exactly
counterbalanced by the internal pressure of the hot gas and by the pressure of radiation. Therefore, the
static-equilibrium equation is
\begin{displaymath}
\frac{GM^2}{6RV}=\mathrm{k_B} T n+\frac{\pi^2}{15}\frac{(\mathrm{k_B} T)^4}{(\hbar\mathrm{c})^3}\; ,
\end{displaymath}
with $V\propto R^3$.

If we impose that $n\equiv n_*$, $\mathrm{k_B} T\equiv\mathrm{k_B} T_*$ and $M\equiv M_*\equiv M_{\odot}$, we obtain
the following relation between the radius of a stellar core in equilibrium conditions, $R_*$, and the actual solar
radius, $R_{\odot}$,
\begin{equation}
R_*\simeq 0.21R_{\odot}\; .  \label{radius}
\end{equation}

By comparing the quantities $n$, $T$ and $M$, we have seen, through Eq.~(\ref{radius}), that the core of a star like the
Sun is not in an equilibrium state globally thermodynamically stable, although the equilibrium Maxwellian
distribution, used for solar core description, is a good approximation. The core can be better described as a
metaequilibrium or metastable state. This state may be due to temperature fluctuations~\cite{beck} or 
quasi-particle behavior of the ions of plasma~\cite{CLMQ}.

%%%%%%%%%%%%%%%%%%%%%%%%%%%%%%%%%%%%%%%%%%%%%%%%%%%%%%%%%%%%
\section{The number of particles inside the \debhu\/ sphere}
%%%%%%%%%%%%%%%%%%%%%%%%%%%%%%%%%%%%%%%%%%%%%%%%%%%%%%%%%%%%
Before reporting on the application to main sequence stars (with masses close to the solar mass) of free energy density minimization, we wish to recall the fully analytical descriptions of stellar models and stellar stability by Clayton~\cite{clayton}, Nauenberg and Weisskopf~\cite{nauenberg1}, and Balian and Blaizot~\cite{BB}.

As reported in these papers, the equilibrium properties are obtained by hydrostatic equations with the addition of
local thermodynamical equations for matter and radiation, based on Boltzmann-Gibbs entropy, with the consequence that
stars evolve along a sequence of states in a complete hydrostatic and thermal equilibrium. However, as we shall
verify, in these states physical quantities like density, temperature and core radius of stars with the composition
and size of the Sun are not those required to achieve a global free energy density minimum, that, rather, is a local
minimum with a very long lifetime, because of many-body interactions. Following Sewell, this state is a metastable
state which (as shown for instance by Sakagami and Taruya~\cite{ST}) can more conveniently be described by a
non-extensive distribution function.

The star interior is a hot dense electron-nuclear plasma at high pressure and temperature. Density and temperature are
growing toward the center, because we require that the system be stable in the gravitational field. At zero
approximation, plasma core can be considered as a Boltzmann ideal gas, with energy $3N\mathrm{k_B} T/2$. SSM, based on ideal gas approximation of the core~\cite{bahcall1,castellani}, correctly evaluates the measured solar neutrino fluxes. SSM cannot account for the existence of density perturbations since it is based on hydrostatic evolution equation.

In the hydrodynamical approximation, density perturbations can be induced by corresponding temperature fluctuations due
to convection of matter between layers with different local temperatures.

Greater is the quantity $(\mathrm{k_B} T)^{3/2}n^{-1/2}$ (where $n$ is, as usual, the number density), 
better approximation of the stellar core is the equilibrium Maxwellian distribution. In fact, this
quantity is related to the square number $N_{\mathrm{D}}$ of particle inside the \debhu\/ sphere, which must be sufficiently large in order to apply Boltzmann-Gibbs statistics.

If we define the volume of the \debhu\/ sphere $V_{\mathrm{D}}$,
\begin{displaymath}
V_{\mathrm{D}}=\frac{4}{3}\pi R_{\mathrm{D}}^3=\frac{4}{3}\pi\left(\frac{\mathrm{k_B} T}{4\pi\e^2\sum_i n_i
Z_i^2}\right)^{3/2},
\end{displaymath}
and the number of particles $N_{\mathrm{D}}$ inside the \debhu\/ sphere,
\begin{displaymath}
N_{\mathrm{D}}=n V_{\mathrm{D}}\propto \frac{(\mathrm{k_B} T)^{3/2}}{n^{1/2}} \; ,
\end{displaymath}
and assuming that $V_{\mathrm{D}}$ contains only protons, helium nuclei and electrons, for a star with composition
similar to that of the solar core, we obtain $N_{\mathrm{D}}\approx 4$. This is a very small number to apply, without
criticism, the Boltzmann-Gibbs statistics, because the fluctuations of that number are great. Furthermore, the value
of $N_{\mathrm{D}}$ can increase if $\mathrm{k_B} T$ increases and/or the density $n$ decreases.

Going on with time (few billion of years) the core of stars evolves toward states that are closer to equilibrium, but
they deviate from this behavior when the helium burning becomes the most important fusion process. At $t\simeq 8\cdot
10^9\,\mathrm{y}$ we have an electron-nuclear (hydrogen) plasma with a density lower than the actual density and a
temperature much higher~\cite{Claytonbook}; therefore this state (pre-helium burning state) is closer to global
equilibrium and the Maxwellian distribution describing this state is a better approximation than at an earlier time
because $(\mathrm{k_B} T)^{3/2}n^{-1/2}$ is greater.

A state with a great value of $(\mathrm{k_B} T)^{3/2}n^{-1/2}$ is closer to equilibrium than a state with a lower
value. Let us also note that, although $R_{\mathrm{D}}\propto(\mathrm{k_B} T/n)^{1/2}$ is constant along the star
profile, $(\mathrm{k_B} T)^{3/2}n^{-1/2}$ is not.

Of course, as $N_D$ increases, the weak electron screening correction
\begin{displaymath}
f^{\mathrm{WES}}\approx \exp\left(\frac{1}{4\pi}\frac{Z_1 Z_2}{3N_{\mathrm{D}}}\right)\; ,
\end{displaymath}
goes to one.

In non-extensive thermostatistics the free energy of an ideal gas (without considering corrections at the moment) is
given by~\cite{plastino}
\begin{displaymath}
F_q=-\mathrm{k_B} T\ln_q\left[\Lambda d(q) V^{N_{\mathrm{D}}} (\mathrm{k_B}T)^{3/2N_{\mathrm{D}}}\right]\; ,
\end{displaymath}
where
\begin{displaymath}
\Lambda=\frac{1}{N_{\mathrm{D}}!}\left(\frac{m}{2\pi\hbar^2}\right)^{3N_{\mathrm{D}}/2},
\end{displaymath}
$\ln_q x=(x^{1-q}-1)/(1-q)$ is the $q$-logarithm and $d(q)$ is a suitable function of the entropic parameter. Let us
apply the above expression of $F_q $ to the \debhu\/ sphere. When $q\neq 1$, in order to have a minimum of $F_q$, one
needs a large value of $R_D$ that means a large value of $(\mathrm{k_B} T)^{3/2}n^{-1/2}$. Of course, as $q\rightarrow
1$ the free energy reaches the minimum for fixed $T$ and $n$.

The entropy of the subsystem can be deduced from the relation
\begin{displaymath}
F=TS-E_{\mathrm{tot}}\; ,
\end{displaymath}
the expression of $F$ being well-known.

Neglecting two terms in comparison to others, we can write
\begin{equation}
\frac{S}{\mathrm{k_B}}\simeq \frac{5}{2}N_{\mathrm{D}}\left[+1+n\frac{e_1}{(\mathrm{k_B}T)^{3/2}}
-n^{1/2}e_2\frac{(\bar{Z}+1)^{3/2}}{(\mathrm{k_B}T)^{3/2}}\right]\; .  \label{Sq first}
\end{equation}

We know that the non-extensive Tsallis entropy can be written, for a system with a small $N_{\mathrm{D}}$,
as~\cite{Quarati&Quarati}
\begin{equation}
\frac{S_q}{\mathrm{k_B}}=\frac{5}{2}N_{\mathrm{D}}\left[+1+\frac{1-q}{2}\frac{5}{2}N_{\mathrm{D}}\right]\; . \label{Sq
second}
\end{equation}

Equalling the two relations in Eqs.~(\ref{Sq first}) and~(\ref{Sq second}) and substituting $n_*$ and $T_*$ with
$N_{\mathrm{D}}\approx 4$, we obtain that in the \debhu\/ sphere the $q$ parameter must assume the value of $%
\approx 1.001$, in order to achieve equilibrium. The correction of one thousand, compared to unity, seems very
negligible. However it can be not at all without importance in the calculation of the fusion rates.

%%%%%%%%%%%%%%%%%%%%%%%%%%%%%%%%%%%%%%%%%%%%%%%%%%%%
\section{Distribution function of metastable states}
%%%%%%%%%%%%%%%%%%%%%%%%%%%%%%%%%%%%%%%%%%%%%%%%%%%%
Non-equilibrium systems with metastable states can be composed of regions that show fluctuations in space and time of an intensive quantity like the inverse temperature $\beta$~\cite{beck}.

Non-equilibrium metastable states of a macroscopic system are made of cells that are temporarily in local equilibrium.
Each small region has a different $\beta$, varying in accordance with the probability distribution function $f(\beta)$. Therefore,
stationary probability density of a non-equilibrium system are given by
\begin{displaymath}
P(\varepsilon)=\int_0^{+\infty}\d\beta f(\beta)\e^{-\beta\varepsilon}\; .
\end{displaymath}

The function $f(\beta)$ can assume several expressions. However, in case of small deviations from Maxwellian
distribution, the $P(\varepsilon)$ function can always be written as~\cite{beck}
\begin{displaymath}
P(\varepsilon)=\e^{-\beta_0\varepsilon}\e^{-\delta(\beta_0\varepsilon)^2}=\e^{-\beta_0\varepsilon}
\left[1+\frac{1}{2}(q-1)\beta_0^2\varepsilon^2+\dots\right]\;,
\end{displaymath}
where $\delta=(1-q)/2$, $q$ is the Tsallis non-extensive parameter, and $\beta_0$ is the average value of the quantity
$\beta$. In this case, if the distribution $f(\beta)$ has a standard deviation $\sigma$, we obtain the useful relation
\begin{displaymath}
\sqrt{q-1}=\frac{\sigma}{\beta_0}\; .
\end{displaymath}

$P(\varepsilon)$ is proportional to non-extensive Tsallis distribution when the entropic parameter $q$ is very close to unity.

For a pure self-gravitating system Sakagami and Taruya~\cite{ST} have shown that, extremizing non-extensive Tsallis
entropy, a stellar polytrope distribution can be obtained. This polytrope function is regarded as a finite isothermal
polytrope, usually as an equilibrium state. However, the one-particle distribution function of stellar polytrope shows velocity spatial dispersion. Therefore it is no longer the
equilibrium but the quasi-equilibrium state (metastable state).

A stellar core with $n$, $T$ and $M$ different from the $n_*$, $T_*$ and $M_*$ equilibrium values can more properly be
described by a non-extensive distribution of a metaequilibrium state, rather than by the Maxwellian equilibrium distribution (that can be seen as a zero approximation).

Temperature fluctuations $\Delta T/T$ of order between $1\%$ and $8\%$ 
have been invoked to analyse neutrino oscillation parameters~\cite{nunokawa,Guzzo}. It is straightforward to show that fluctuations of such an order of magnitude are
related to values of $\delta$ of about $10^{-4}\div 10^{-3}$. With the use of these small deformations, we do not change the macroscopic features of the core and its bulk properties; with the modified nuclear rates, we evaluate neutrino fluxes in agreement with the measured ones and, within
actual uncertainties, these deformations should be taken into consideration for a more precise determination of the oscillation parameters.

This kind of fluctuation could be ruled out in an electron-nuclear plasma, after analysis of experimental 
results and global fits. Fluctuations of other intensive quantities 
can be responsible of metastable states. Fluctuating quantities can be, for instance, effective chemical potentials or functions of fluctuating energy dissipation in the flow (for the turbulence application). 

Other effects that can be responsible of deviations from global thermodynamical equilibrium should be investigated, as reported below.
 
Li and Zhang~\cite{liz} have discussed how solar core may be far from equilibrium, but rather in a 
quasi-stationary state, investigating the self-generated magnetic fields in the solar core~\cite{li}. 
In thermal nuclear fusion plasmas, where inelastic ions collisions (e.g. $pp$ chain) produce thermal energy, 
dissipative effects of thermal Bremsstrahlung-radiation loss-emission cause a stellar core state 
which differs from a global equilibrium state. Elastic collisions frequency of the order of $10^{17}$ Hz would destroy 
collective motion and coherent structure in the plasma if the solar system were isolated adiabatically. 
In the case of an isolated system, the Sun would be in thermodynamical equilibrium. However, the solar system is an 
open system and the nonequilibrium effect cancels the collision effect. Consequences are that nonequilibrium can be a 
source of order, the existence of random radiation fields (random electric microfields distribution and magnetic field 
distribution) are due to self-organization phenomena. For a discussion of the r\^{o}le 
of random electric microfield component see Refs.~\cite{LavQua,ka1,ka2}; see also Jiulin Du \cite{jiulin} for 
a discussion on self-gravitation systems.

Grandpierre~\cite{grand} has discussed how the core is neither a ball of ideal gas, nor a quiescent 
steady-state fusion reactor, but a complex self-organizing system. 

Burgess {\it et al.}~\cite{burges}, among others, 
report on the importance of the resonance between g-modes and magnetic Alfv\'en 
waves in the solar radiative zone, how strongly density variations affect the solar neutrino survival probability and 
how oscillation parameters depend on the magnitude of solar density fluctuations.

The presence of at least one of these effects can influence the statistical distribution function and be the cause 
of deviations from Maxwellian distribution. We argue that these effects can be taken into account by the 
appropriate value of the entropic parameter $q$~\cite{beck2}.

%%%%%%%%%%%%%%%%%%%%%
\section{Conclusions}
%%%%%%%%%%%%%%%%%%%%%
The system we have considered in this work is the electron-nuclear plasma of the core of Sun-like stars. We can deduce
that the larger is the magnitude of the quantity $(\mathrm{k_{B}}T)^{3}n^{-1/2}$, the nearer the system is to
the globally thermodynamically stable equilibrium state, because the number of particles inside the \debhu\/ sphere
increases and reaches its maximum as $(\mathrm{k_{B}}T)^{3}n^{-1/2}$ has the maximum value. In this case, the use of
the Boltzmann-Gibbs statistics is fully justified and the free energy has a minimum; also the free energy of
non-extensive statistics has a minimum.

We have looked for the minimum of free energy density as stated by Sewell to find the global equilibrium states.
Considering the corrections to an ideal gas due to identity of particles and to inter-nuclear interaction and the
black-body radiation emitted, by minimizing the free energy density of the electron-nuclear plasma, we have obtained
the following values
\begin{displaymath}
n_{\ast }\simeq 2.74\cdot 10^{-14}\,\mathrm{fm}^{-3}\;\;,\;\;\mathrm{k_{B}} T_{\ast }\simeq 5\cdot
10^{-3}\,\mathrm{MeV}\;\;\mathrm{and}\;\;R_{\ast }\approx 0.2R_{\odot }\;,
\end{displaymath}
with a typical chemical composition $\bar{Z}=1.25$.

States with different values of $\mathrm{k_{B}}T$ (lower) and $n$ (higher) are metastable states that can be featured
by temperature fluctuations or density fluctuations, by quasi-particle models or by the presence of self-generated magnetic fields or random microfields distributions. These physical effects are
responsible of metastable distributions that are given by expressions that may be derived from the non-extensive
thermostatistics.

Deviations from momentum Maxwellian distribution are quite small for the astrophysical systems considered in this paper and the
main properties of the stars do not change at all for these deviations. Nevertheless, very slight deviations 
can sensibly affect the evaluation of nuclear fusion rates (as we have shown elsewhere both
for nuclear astrophysics processes and for heavy ion physics and quark-gluon plasma properties) 
and could be useful in solving the problem of lithium abundance in the universe, among many other applications.

\vspace{1cm}
\noindent
{\it Supported in part by MIUR-PRIN2003.}

%%%%%%%%%%%%%%%%%%%%%%%%%%%%%%%%%%%%%%%%%%%%%%%%%%%%%%%%%%%%%%%%%%%%%%%%%%%%%%%%%%%%%%%%%%%%%%%%%%%%%%%%%%%%%%%%
%%%%%%%%%%%%%%%%%%%%%%%%%%%%%%%%%%%%%%%%%%%%%%%%%%%%%%%%%%%%%%%%%%%%%%%%%%%%%%%%%%%%%%%%%%%%%%%%%%%%%%%%%%%%%%%%

\end{document}